\documentclass[preprint,noshowpacs,noshowkeys,preprintnumbers,amsmath,amssymb]{revtex4}
\usepackage{graphicx}
\usepackage{dcolumn}
\usepackage{bm,braket}

\DeclareMathOperator{\diag}{diag}

\DeclareMathOperator{\Img}{Im}
\DeclareMathOperator{\Tr}{Tr}
\newcommand{\rnum}[1]{\expandafter{\romannumeral#1}}
\newcommand{\Rnum}[1]{\uppercase\expandafter{\romannumeral#1}}
\newcommand{\expect}[1]{\langle#1\rangle}
\newcommand{\abs}[1]{\left\lvert#1\right\rvert}
\newcommand{\trans}[1]{{#1}^\mathsf{T}}

\begin{document}
\preprint{}
\title{Thermal leptogenesis scenarios \\in the restrictive left-right symmetric model}
\author{Yuya Wakabayashi}
\affiliation{Department of Physics, Rikkyo University, Tokyo 171-8501, Japan}
\email{waka@stu.rikkyo.ne.jp}

\begin{abstract}
We investigated thermal leptogenesis scenarios in the left-right symmetric extension of the standard model. Imposing the $D$-parity realization below GUT scale and the grand unification make our model more restrictive and predictive. In such a case, a $D$-parity odd singlet has a critical role. This singlet have prospects of causing a very large mass hierarchy between $\mathit{SU}(2)_{L,R}$ triplet scalars.

We test our model by computing baryogenesis via leptogenesis. Our model has two sources of the lepton number asymmetry in the universe, the heavy right-handed neutrinos $N_i$ and the $\mathit{SU}(2)_L$ triplet scalar $\Delta_L$. Leptogenesis scenarios can be categorized by these mass scales. If the light neutrinos are Majorana and have a hierarchical mass spectrum, we can obtain a successful result in leptogenesis through $N_1$-decay. But we found that the normal mass hierarchy of the light neutrinos can conflict with leptogenesis through $\Delta_L$-decay in the SM. In order to obtain the successful thermal leptogenesis through $\Delta_L$-decay, we need to introduce more Higgs doublets. This result suggest the two Higgs doublet model with an $\mathit{SU}(2)_L$ triplet scalar.
\end{abstract}

\maketitle
\section{Introduction}
The $\mathit{SO}(10)$ gauge theory is a very attractive candidate of the grand unification. First feature is the unification of three gauge interactions. This feature can rewrite the Gell-Mann--Nishijima relation to $Q = T_L^3+T_R^3+(B-L)/2$, i.e. the electric charge can be quantized and related to our classically familiar charges, the baryon number $B$ and the lepton number $L$. Second fascinating feature is the matter unification. Differently from the $\mathit{SU}(5)$ GUT, each quarks and leptons corresponds to a 5-bit eigenstate of Cartan subalgebra:
\begin{equation}
\begin{aligned}
\nu_{eL} &= \ket{\downarrow\downarrow\downarrow;\uparrow\downarrow},&
u_L^r &= \ket{\downarrow\uparrow\uparrow;\uparrow\downarrow},&
u_L^b &= \ket{\uparrow\downarrow\uparrow;\uparrow\downarrow},&
u_L^g &= \ket{\uparrow\uparrow\downarrow;\uparrow\downarrow},\\
e_L^- &= \ket{\downarrow\downarrow\downarrow;\downarrow\uparrow},&
d_L^r &= \ket{\downarrow\uparrow\uparrow;\downarrow\uparrow},&
d_L^b &= \ket{\uparrow\downarrow\uparrow;\downarrow\uparrow},&
d_L^g &= \ket{\uparrow\uparrow\downarrow;\downarrow\uparrow},\\
-(\nu_{eL})^c &= \ket{\uparrow\uparrow\uparrow;\downarrow\uparrow},&
-(u_L^r)^c &= \ket{\uparrow\downarrow\downarrow;\downarrow\uparrow},&
-(u_L^b)^c &= \ket{\downarrow\uparrow\downarrow;\downarrow\uparrow},&
-(u_L^g)^c &= \ket{\downarrow\downarrow\uparrow;\downarrow\uparrow},\\
e_L^+ &= \ket{\uparrow\uparrow\uparrow;\uparrow\downarrow},&
(d_L^r)^c &= \ket{\uparrow\downarrow\downarrow;\uparrow\downarrow},&
(d_L^b)^c &= \ket{\downarrow\uparrow\downarrow;\uparrow\downarrow},&
(d_L^g)^c &= \ket{\downarrow\downarrow\uparrow;\uparrow\downarrow},&
\end{aligned}
\label{16matters}
\end{equation}
here $\uparrow$ and $\downarrow$ in the kets denotes the eigenvalues of five generators of Cartan subalgebra: First three arrows for $\mathit{SO}(6) \simeq \mathit{SU}(4)_c$, and last two for $\mathit{SO}(4) \simeq \mathit{SU}(2)_L\!\times\!\mathit{SU}(2)_R$. Consequently, by adding three right-handed neutrinos, we can obtain a economic picture of matter in  the universe.

Unfortunately, however, the GUT scale physics is often highly surpressed. Then we focus on the left-right symmetric extension of the standard model (SM) as the low-energy effective theory of the $\mathit{SO}(10)$ theory. The left-right symmetric model (LR) \cite{LR} is one of the oldest extension of particle physics. Since the LR model, like the $\mathit{SO}(10)$, naturally introduce right-handed neutrinos, and then it harmonize well with the recent evidences of the nonzero neutrino masses. Generally, this model requires the $\mathit{SU}(2)$ triplet scalars in order to reproduce the tiny neutrino masses. These triplets could become a smoking-gun evidence of the LR model. The LR models in the literature have more free parameters and must be less predictive. We can obtain more restrictive scenarios by considering the LR with the $D$-parity and the grand unification.

In order to approach the ultra high energy scale, we consider baryogenesis via leptogenesis in the universe \cite{leptogenesis1}. The WMAP collaboration showed the baryon-to-photon number ratio with high precision \cite{WMAP}:
\begin{equation}
\eta_B^\text{CMB} \equiv \frac{n_B}{n_\gamma} = \frac{n_b-n_{\overline{b}}}{n_\gamma} = (6.14\pm0.25)\times10^{-10}.
\end{equation}
Also second determination of $\eta_B$ can be obtained from nucleosynthesis, i.e. abundances of the light elements, $\mathrm{D}$, $\mathrm{{}^3He}$, $\mathrm{{}^4He}$, $\mathrm{{}^7Li}$ \cite{BBN}:
\begin{equation}
\eta_B^\text{BBN} = (2.6\text{--}6.2)\times10^{-10}.
\end{equation}
The full $\mathit{SO}(10)$ GUT framework has been extensively studied \cite{SO(10)}. We analyze whether the observed values of $\eta_B = \mathcal{O}(10^{-10})$ can be reproduced in order to investigate our restrictive and minimal LR model.

This paper is arranged as follows. In Sec.~\ref{ModelDescription}, we briey review the LR extended model and setup our model. As remarked above, in order to make our model more restrictive and predictive, we impose the D-parity restoration and the some grand unication like $\mathit{SO}(10)$. Here the GUT means that all the gauge couplings come together as one at high energy scale (Sec.~\ref{NumericalAnsatz}) and the unification of the known quark and lepton fields (Sec.~\ref{NeutrinoSpectrum}). Keep in mind that this GUT condition does not necessarily imply the $\mathit{SO}(10)$ GUT. There we show that a hierarchical structure of the GUT scale and the $D$-parity restoration scale is essential. Sec.~\ref{NeutrinoSpectrum} provides the neutrino mass spectrum. We succeed in obtaining the hierarchical heavy neutrino spectrum by means of the normal hierarchy of light neutrinos and a few assumptions. We discuss baryogenesis via leptogenesis for some distinct scenarios in Sec.~\ref{Baryogenesis}.

\section{Models}
\subsection{Building more constrained models}\label{ModelDescription}
We consider the left-right symmetric breakdown of the grand unification. A simplest candidate of the grand unification of this type is $\mathit{SO}(10)$ and their minimal sets of the Higgs multiplets are listed below:
\begin{itemize}
\item{$\bm{\textbf{GUT} \to \underline{G_{3221}\!\times\!\mathit{D} \left(\to G_{3221}\right) \to G_{321}}}$}\\
This requires a set of scalars $\bm{210}$, $\bm{126}\oplus\overline{\bm{126}}$ and $\bm{10}$ of $\mathit{SO}(10)$ group. An $\mathit{SU}(4)_c$ adjoint representation $(\bm{15},\bm{1},\bm{1}) \in \bm{210}$ breaks $\mathit{SO}(10)$.
\item{$\bm{\textbf{GUT} \to G_{422}\!\times\!\mathit{D} \left(\to G_{422}\right)\to G_{3221} \to G_{321}}$}\\
This scenario have a need for $\bm{54}$, $\bm{45}$, $\bm{126}\oplus\overline{\bm{126}}$ and $\bm{10}$. An $\mathit{SU}(4)_c$ adjoint representation $(\bm{15},\bm{1},\bm{1}) \in \bm{45}$ breaks $G_{422}$.
\item{$\bm{\textbf{GUT} \to G_{422}\!\times\!\mathit{D} \left(\to G_{422}\right) \to G_{421} \to G_{321}}$}\\
Although in this breaking chain an essential set of scalars is as same as the above, i.e. $\bm{54}$, $\bm{45}$, $\bm{126}\oplus\overline{\bm{126}}$ and $\bm{10}$, an $\mathit{SU}(2)_R$ adjoint representation $(\bm{1},\bm{1},\bm{3}) \in \bm{45}$ partially breaks the right-handed isospin symmetry.
\end{itemize}
here $G_{3221}$, $G_{422}$ and $G_{321}$ denotes the LR gauge group $\mathit{SU}(3)_c\!\times\!\mathit{SU}(2)_L\!\times\!\mathit{SU}(2)_R\!\times\!\mathit{U}(1)_{B{-}L}$, the Pati-Salam (PS) group $\mathit{SU}(4)_c\!\times\!\mathit{SU}(2)_L\!\times\!\mathit{SU}(2)_R$ and the SM gauge group $\mathit{SU}(3)_c\!\times\!\mathit{SU}(2)_L\!\times\!\mathit{U}(1)_Y$ respectively. And the numbers in the brakets denotes the PS quantum numbers. In order to obtain more predictive models we take the Michel's conjecture into consideration. Generally in $\mathit{SO}(10)$ grand unification, quarks and leptons are asigned into three $16$-dimensional spinor representations as listed in Eq.~\eqref{16matters}. And the gauge fields are in a $45$-dimensional adjoint representation.

It should be noted that the PS-singlet in $\bm{210}$ is axial under the $\mathit{D}$-parity, howver, the one in $\bm{54}$ is not. This difference occupies an important place in our model building. Furthermore, since first candidate have the least breaking steps, it would be expected to be the most constrained case. Then we consider the minimal LR model and use the GUT realization as a boundary condition at higher energy scale. As a result, we find that it is sure as expected. We show the details in the next subsection.

\subsection{Numerical ansatz}\label{NumericalAnsatz}
At first, we show all required scalar representations underlined in the above:
\begin{align}
\sigma &= (\bm{1},\bm{1},\bm{1},0) \in \bm{210},\\
\Delta_L &= (\bm{1},\bm{3},\bm{1},+2) = \begin{pmatrix}\Delta_L^+/\sqrt2&\Delta_L^{++}\\\Delta_L^0&-\Delta_L^{++}/\sqrt2\end{pmatrix} \in \bm{126},\\
\Delta_R &= (\bm{1},\bm{1},\bm{3},+2) = \begin{pmatrix}\Delta_R^+/\sqrt2&\Delta_R^{++}\\\Delta_R^0&-\Delta_R^{++}/\sqrt2\end{pmatrix} \in \bm{126},\\
\Phi &= (\bm{1},\bm{2},\bm{2},0) \in \bm{10},
\end{align}
here we showed the $G_{3221}$ quantum numbers. The bidoublet $\Phi$ corresponds to the two Higgs doublets. Hereafter we denote the SM and the extra ones as $H$ and $H'$ respectively. This model is known as the minimal LR model with the spontaneously broken $\mathit{D}$-parity. This model is depicted as
\begin{equation}
G_{3221}\!\times\!\mathit{D} \xrightarrow{\expect{\sigma}} G_{3221} \xrightarrow{\expect{\Delta_R}} G_{321} \xrightarrow{\expect{\Phi}} \text{QCD}\!\times\!\text{QED},\label{BreakingChain}
\end{equation}
and the required vev's are given by \cite{LR}
\begin{align}
\expect{\Delta_L} &= \begin{pmatrix}0&0\\v_L&0\end{pmatrix},&
\expect{\Delta_R} &= \begin{pmatrix}0&0\\v_R&0\end{pmatrix},&
\expect{\Phi} &= \begin{pmatrix}\kappa_1&0\\0&\kappa_2\end{pmatrix}.
\end{align}
Here we ignored the relative phase between $\kappa_1$ and $\kappa_2$. Phenomenologically $v_R \gg \kappa_+ \gg v_L$ is required, where $\kappa_+$ is the standard electroweak breaking vev, $\kappa_+^2 \equiv \kappa_1^2+\kappa_2^2$. In this model, according to the extreme value analysis of the Higgs potential, we have the following relations~\cite{CMP,Narendra}:
\begin{equation}
v_L \sim -\frac{\beta\kappa_+^2v_R}{2M\eta_\text{P}} \label{vL}
\end{equation}
and \cite{Chang,Narendra}
\begin{align}
M_{\Delta_L}^2 &= \mu_\Delta^2-(M\eta_\text{P}+\gamma\eta_\text{P}^2),\label{MDL}\\
M_{\Delta_R}^2 &= \mu_\Delta^2+(M\eta_\text{P}-\gamma\eta_\text{P}^2),\label{MDR}
\end{align}
where each couplings are defined as follows \cite{LR,vevseesaw}:
\begin{subequations}
\begin{equation}
\begin{aligned}
V
&\ni
M\sigma\left[\Tr\left(\Delta_L\Delta_L^\dagger\right)-\Tr\left(\Delta_R\Delta_R^\dagger\right)\right]
+\gamma\sigma^2\left[\Tr\left(\Delta_L\Delta_L^\dagger\right)+\Tr\left(\Delta_R\Delta_R^\dagger\right)\right]\\
&\quad
+\beta_1\left[\Tr\left(\Phi\Delta_R\Phi^\dagger\Delta_L^\dagger\right)+\Tr\left(\Phi^\dagger\Delta_L\Phi\Delta_R^\dagger\right)\right]
+\beta_2\left[\Tr\left(\tilde{\Phi}\Delta_R\Phi^\dagger\Delta_L^\dagger\right)+\Tr\left(\tilde{\Phi}^\dagger\Delta_L\Phi\Delta_R^\dagger\right)\right]\\
&\quad +\beta_3\left[\Tr\left(\Phi\Delta_R\tilde{\Phi}^\dagger\Delta_L^\dagger\right)+\Tr\left(\Phi^\dagger\Delta_L\tilde{\Phi}\Delta_R\dagger\right)\right]
+\beta_4\left[\Tr\left(\tilde{\Phi}\Delta_R\tilde{\Phi}^\dagger\Delta_L^\dagger\right)+\Tr\left(\tilde{\Phi}^\dagger\Delta_L\tilde{\Phi}\Delta_R^\dagger\right)\right].
\end{aligned}
\end{equation}
\begin{equation}
\beta_{ab} =
\begin{pmatrix}
\beta_1 & \beta_2\\
\beta_3 & \beta_4
\end{pmatrix}
\end{equation}
\end{subequations}
And $\eta_\text{P}$ denotes the vacuum expectation value of $\sigma$ field potential, which is defined by
\begin{align}
\eta_\text{P} &\equiv \expect{\sigma} = \frac{\mu_\sigma}{\sqrt{2\lambda_\sigma}},&
V_\sigma &= -\mu_\sigma^2\sigma^2+\lambda_\sigma\sigma^4.
\end{align}
The relation like Eq.~\eqref{vL} is known as the vev seesaw mechanism \cite{vevseesaw}. We see that Eq.~\eqref{MDL} and \eqref{MDR} are not symmetric in spite of the left-right symmetric framework. This results from the fact that the scalar $\sigma$ is an axial under the $\mathit{D}$-parity. We ontain two important indications from these relations. First, according to Eq.~\eqref{vL}, the hugeness of the $\mathit{D}$-parity breaking scale $\eta_\text{P}$ results in the smallness of the vev of $\Delta_L$~\cite{Narendra}. We numerically check this fact later. Second, let us consider Eq.~\eqref{MDL} and \eqref{MDR}. If $M\eta_\text{P}$ and $\gamma\eta_\text{P}^2$ are at the same order, they are canceled each other and then we have $M_{\Delta_R}^2 \sim \mu_\Delta^2$. If such is the case, the squared mass of $\Delta_L$ would be given by $M_{\Delta_R}^2-2\gamma\eta_\text{P}^2$. Hence it would be possible that the left-handed triplet $\Delta_L$ is much lighter than the right-handed one $\Delta_R$:
\begin{equation}
M_{\Delta_L}^2 \ll M_{\Delta_R}^2. \label{hierarchy}
\end{equation}
Below we take this possibility into account.

Since it is absolutely imperative for leptogenesis to generate the lepton asymmetry, we are interested in when the $\mathit{SU}(2)_R\!\times\!\mathit{U}(1)_{B{-}L}$ symmetry broke down. For this end we solve the two-loop renormalization equations for the gauge couplings. As remarked before, in order to obtain more restrictive models we  impose two boundary conditions. First boundary condition is the restoration of the $\mathit{D}$-parity. This means that above a high energy scale the $\mathit{SU}(2)_R$ gauge coupling evolves along with the $\mathit{SU}(2)_L$ one.  Second boundary condition is the grand unification, which is essential to refer to $\mathit{U}(1)$ gauge groups. This constraint suggests that the hypercharge and $B{-}L$ charge are normalized as follows:
\begin{align}
\tilde{Y} &= \sqrt{\frac{3}{5}}Y,&
\tilde{V} &= \sqrt{\frac{3}{2}}\frac{B{-}L}{2}.
\end{align}
And then at the GUT scale, we have
\begin{equation}
\alpha_s(M_\text{GUT}) = \alpha_L(M_\text{GUT}) = \alpha_R(M_\text{GUT}) = \alpha_V(M_\text{GUT}).
\end{equation}
Before showing the results, let us refer  the numbers of the free input parameters. In this renormalization group analysis, we have four input parameters $M_{Z_R}$, $\theta_R$, $m_{H'}$ and $M_{\Delta_L}$. Here $m_{H'}$ denotes the mass of the second doublet Higgs, and the mixing angle $\theta_R$ is defined by \cite{mixture}
\begin{equation}
\begin{pmatrix}B^0_R\\Z^0_R\end{pmatrix} =
\begin{pmatrix}\cos\theta_R&\sin\theta_R\\-\sin\theta_R&\cos\theta_R\end{pmatrix}
\begin{pmatrix}B'^0\\W^0_R\end{pmatrix}.
\end{equation}
We assume that $\Delta_R$ lives in the same energy scale as $Z_R$, and  then we do not treat $M_{\Delta_R}$ as a free input parameter. At this stage the high-precision measurement of $\alpha_2(M_W)$ helps us a lot. Since the current observation error of the electroweak gauge coupling is very small, the allowed range of the mixing angle $\theta_R$ of the neutral $\mathit{SU}(2)_R$ boson $W^0_R$ and $\mathit{U}(1)_{B{-}L}$ boson $B'^0$ is extremely-narrow. Consequently, we can virtually determine $\cos\theta_R$ uniquely, and then we can obtain a strong constraint on the $Z'^0 (\simeq Z^0_R)$ boson mass. In contrast with the usual left-right symmetric models, the consequence of this argument is that a set of three parameters $(M_{Z_R}, m_{H'}, M_{\Delta_L})$ tells us unique values of the $\mathit{SU}(2)_L\!\times\!\mathit{U}(1)_{B{-}L}$ breaking scale $v_R$, the $\mathit{D}$-parity scale $\eta_\text{P}$ and the GUT scale $M_\text{GUT}$. $v_R$ can be calculated from $\alpha_2(Q)$, $\alpha_Y(Q)$, $M_{Z_R}$ and $\theta_R$ through the Newton--Raphson-like method. Note that we can exclude the regions of $M_{\Delta_L} > M_{Z_R}(=M_{\Delta_R})$ and $M_{\Delta_L} < 100$ [GeV]. The former comes from Eq.~\eqref{hierarchy}, an the latter is concluded from the experimental facts. As a direct consequence of Eq.~\eqref{hierarchy}, two triplets $\Delta_L$ and $\Delta_R$ do not mix each other. Then, note that the experimentally detected doubly-charged Higgs boson $h^{++}$ can be identified as pure $\Delta_L^{++}$. The dilepton detection puts the limits of the mass of a doubly-charged Higgs boson $h^{++}$. The decay modes of $\mu\mu$, $ee$ and $e\mu$ bring the results of $M_{h^{++}} > 136, 133, 115$ [GeV] respectively. Also the long-lived doubly-charged Higgs search experiments result in $M_{h^{++}} > 134$ [GeV]. Then we can exclude the latter region.

\begin{figure}
\includegraphics[width=.8\textwidth]{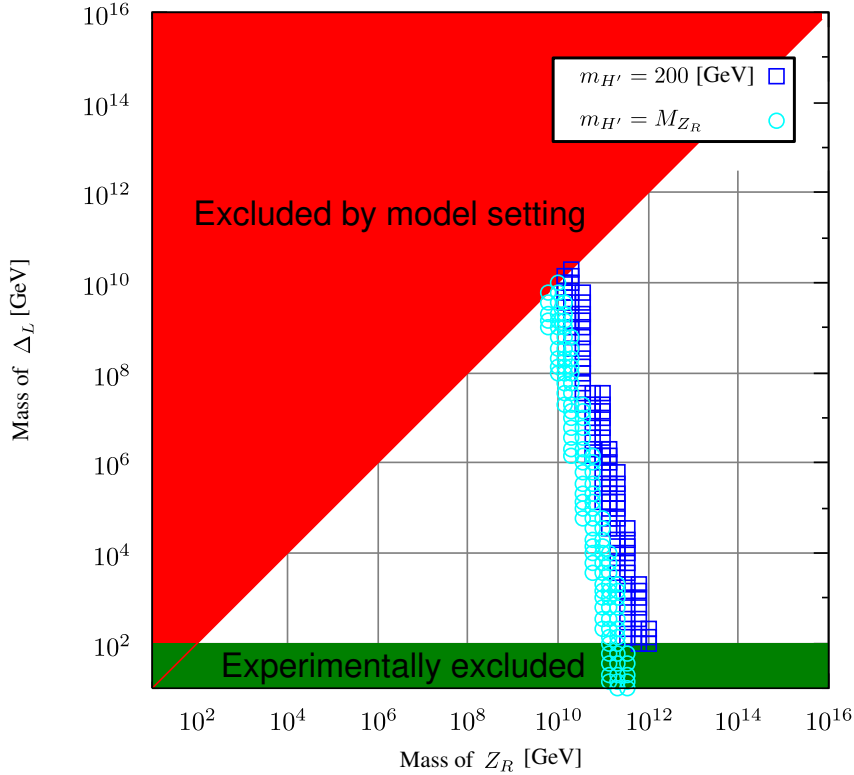} 
\caption{The allowed regions for the grand unifications.}
\label{GUTs}
\end{figure}

Now we show the solutions of the renormalization equations in Fig.~\ref{GUTs}. Here we assume $m_{H'}$ to be $200$GeV or $M_{Z_R}$. We find that in either case $10^{10} \lesssim M_{Z_R} \lesssim 10^{12}$ [GeV], on the other hand, we regret to find that our model setting occurs over a wide range of $M_{\Delta_L}$.

\begin{figure}[htbp]
\begin{minipage}{0.49\textwidth}
\begin{center}
\includegraphics[width=\textwidth]{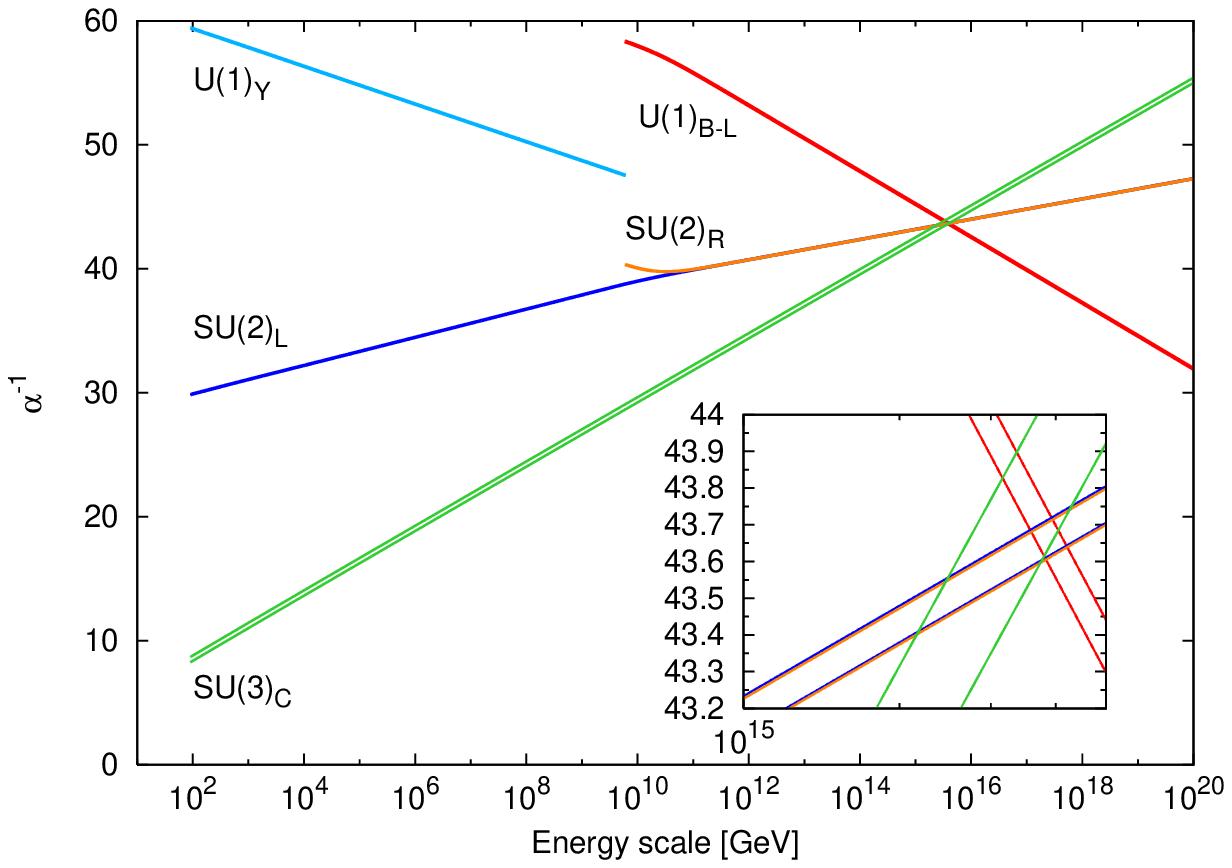}
\caption{\def\baselinestretch{.7}\normalsize SM. \\$m_{H'} = M_{Z_R}$ and $M_{\Delta_L} = 6.0\!\times\!10^9$ [GeV].}
\label{RGE1}
\end{center}
\end{minipage}
\begin{minipage}{0.49\textwidth}
\begin{center}
\includegraphics[width=\textwidth]{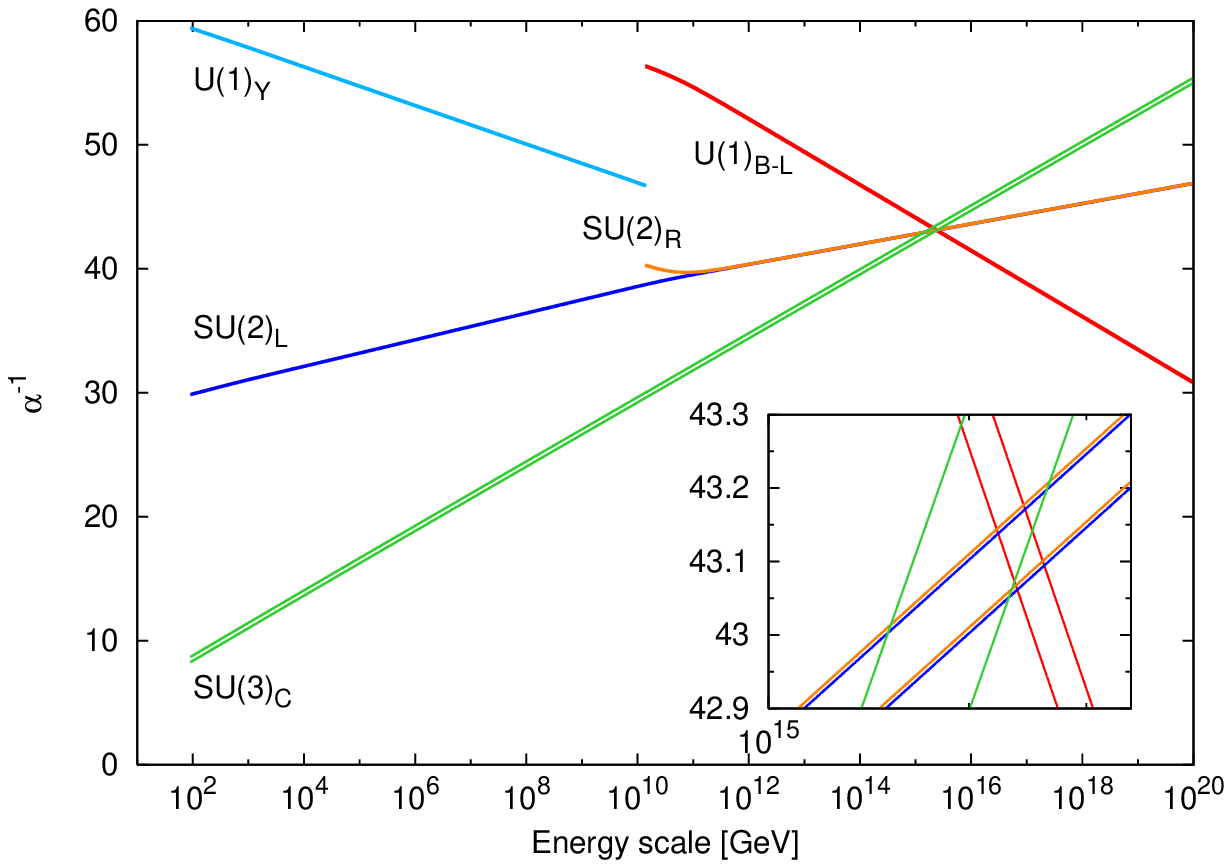}
\caption{\def\baselinestretch{.7}\normalsize 2HDM.\\$m_{H'} = 200$ [GeV] and $M_{\Delta_L} = 1.0\!\times\!10^{10}$ [GeV].}
\label{RGE2}
\end{center}
\end{minipage}
\begin{minipage}{0.49\textwidth}
\begin{center}
\includegraphics[width=\textwidth]{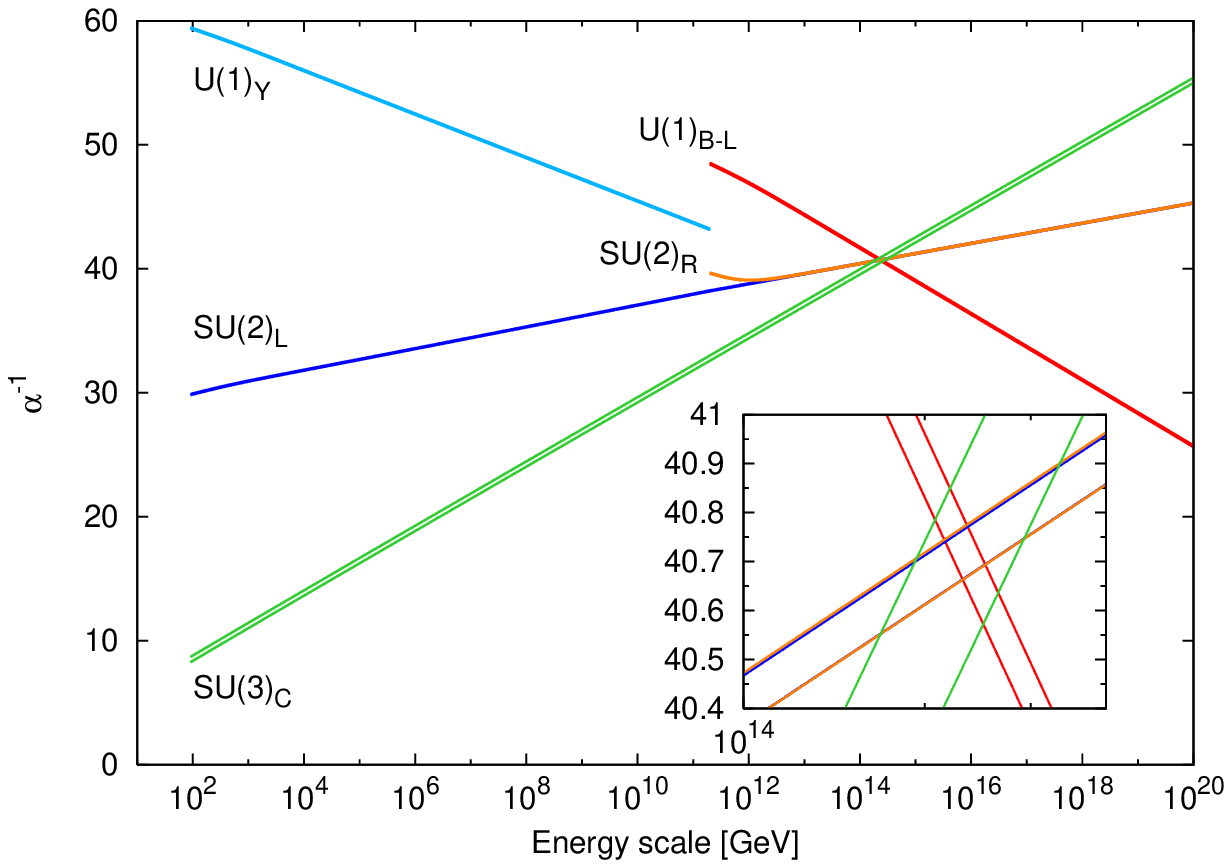}
\caption{\def\baselinestretch{.7}\normalsize SM with an $\mathit{SU}(2)_L$ triplet.\\$m_{H'} = M_{Z_R}$ and $M_{\Delta_L} = 100$ [GeV].}
\label{RGE3}
\end{center}
\end{minipage}
\begin{minipage}{0.49\textwidth}
\begin{center}
\includegraphics[width=\textwidth]{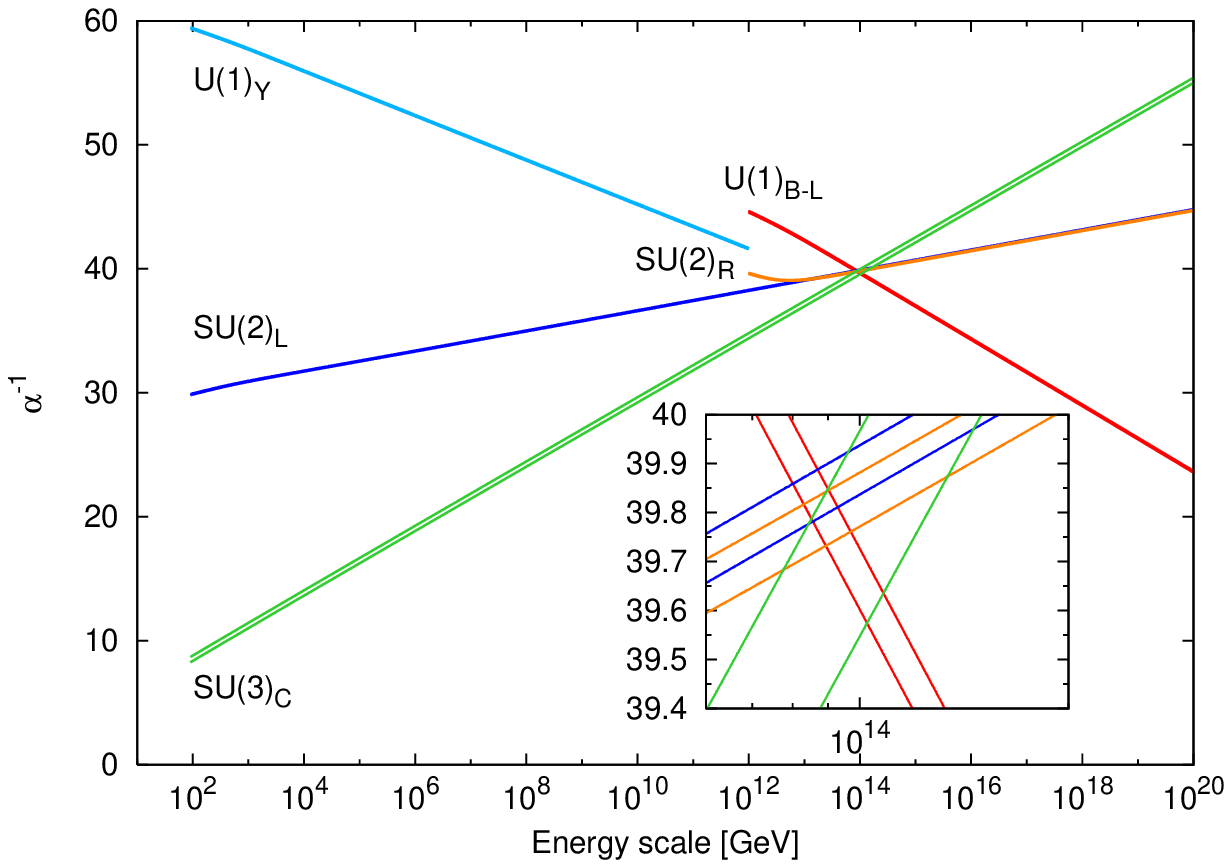}
\caption{\def\baselinestretch{.7}\normalsize 2HDM with an $\mathit{SU}(2)_L$ triplet.\\$m_{H'} = 200$ [GeV] and $M_{\Delta_L} = 100$ [GeV].}
\label{RGE4}
\end{center}
\end{minipage}
\end{figure}

Here let us describe with some illustrations. First we consider the heavier $\mathit{SU}(2)_L$ triplet. We start with Fig.~\ref{RGE1}, where we set $m_{H'} = M_{Z_R}$ and $M_{\Delta_L} = 6\times10^9$ [GeV]. In this case, the low-energy effective theory is the standard model (SM). Fig.~\ref{RGE1} tells us that $v_R = 8.3\times10^9$ [GeV] and $\eta_\text{P} \gtrsim 10^{11}$ [GeV]. Next we refer to Fig.~\ref{RGE2}, which can be obtained by $m_{H'} = 200$ [GeV] and $M_{\Delta_L} = 1.0\times10^{10}$ [GeV]. This case leads the two Higgs doublet model (2HDM) at the low energy scale. We obtain $v_R = 1.9\times10^{10}$ [GeV] and $\eta_\text{P} \gtrsim 10^{11}$ [GeV]. These two parametrizations locate close to the upper ends of Fig.~\ref{GUTs}.

Next we consider the cases of the lighter triplet. These are located close to the lower ends of Fig.~\ref{GUTs}. First, we show the case of $m_{H'} = M_{Z_R}$ and $M_{\Delta_L} = 100$ [GeV] in Fig.~\ref{RGE3}. We obtain $v_R = 2.6\times10^{11}$ [GeV] and $\eta_\text{P} \gtrsim 10^{12}$ [GeV]. And then at the low energy scale, we have the SM with an $\mathit{SU}(2)_L$ triplet. Last, let us consider that both $H'$ and $\Delta_L$ live near the electroweak scale. We show the case of $m_{H'} = 200$ [GeV] and $M_{\Delta_L} = 100$ [GeV] in Fig.~\ref{RGE4}. We obtain $v_R = 1.3\times10^{12}$ [GeV] and $\eta_\text{P} \gtrsim 10^{13}$ [GeV], and we find that the low-energy theory becomes the 2HDM with an $\mathit{SU}(2)_L$ triplet.

Now we give an important comment on these results. The commonly-observed feature is that the computed $\mathit{D}$-parity breaking scales $\eta_\text{P}$'s are much higher than the obtained $\mathit{SU}(2)\!\times\!\mathit{U}(1)_{B{-}L}$ breaking scales $v_R$'s. We find that $\eta_{P}/v_R$'s are larger than $\mathcal{O}(10)$ at the lowest estimate. This is mainly attributable to the mass threshold effects at $v_R$. Substituting this observation to Eq.~\eqref{vL} gives the small value of $v_L$. We use this argument later. Then we can deal with the obtained calculations as guidelines of considering leptogenesis scenarios.

\subsection{Neutrino mass spectrum}\label{NeutrinoSpectrum}
Before solving the sets of Boltzmann equations for leptogenesis, we need the information on the neutrino mass spectrum. Generally the leptonic Yukawa coupling is given by
\begin{equation}
\begin{aligned}
\mathcal{L}_\text{Yukawa}
&= Y_{ij}\overline{\ell_{Li}}\ell_{Rj}\Phi+\tilde{Y}_{ij}\overline{\ell_{Li}}\ell_{Rj}\tilde{\Phi}+\text{h.c}\\
&\quad +{Y_{\Delta}}_{ij}\left[\overline{(\ell_{Li})^c}\ell_{Lj}\Delta_L+\overline{(\ell_{Ri})^c}\ell_{Rj}\Delta_R\right]+\text{h.c}
\end{aligned}
\end{equation}
The left-right stmmetry shows us that the Dirac--Yukawa coupling $Y$ and $\tilde{Y}$ and the Majorana-Yukawa coupling $Y_\Delta$ are hermitian and symmetric in family space respectively: $Y_{ij} = Y_{ij}^\dagger$, $\tilde{Y}_{ij} = \tilde{Y}_{ij}^\dagger$ and ${Y_\Delta}_{ij} = {Y_\Delta}_{ji}$. The symmetry breaking \eqref{BreakingChain} results in the mass matrix of the light neutrinos to be
\begin{align}
m_\nu &= m_\nu^\text{\Rnum{2}}+m_\nu^\text{\Rnum{1}} = Y_\Delta v_L-M_\text{D}{M_R}^{-1}\trans{M_\text{D}},&
{M_R}^{-1} &= \frac{{Y_\Delta}^{-1}}{v_R}.
\label{EffectiveMass}
\end{align}
The generation of the mass suppressed by huge $M_R$ is called as type-\Rnum{1} seesaw mechanism \cite{Seesaw1}, while the mass originated from tiny $v_L$ is the type-\Rnum{2} seesaw mass. As remarked in the last subsection, in all cases we find that $\eta_\text{P}/v_R > \mathcal{O}(10)$. This result rationalize assuming the hierarchical structure of $v_L \ll \max(\kappa_1,\kappa_2) \ll v_R \ll \eta_\text{P}$ (see Eq.~\eqref{vL}). Let us consider that the type-\Rnum{1} mass dominates in the effective neutrino mass matrix \eqref{EffectiveMass}:
\begin{equation}
m_\nu \simeq m_\nu^\text{\Rnum{1}} = -M_\text{D}{M_R}^{-1}\trans{M_\text{D}}.
\label{type1}
\end{equation}
The light neutrino mass matrix $m_\nu$ can be written as
\begin{equation}
m_\nu = U^\ast m_\nu^{\diag} U^\dagger,
\end{equation}
where $U$ denotes the light neutrino mixing matrix, which is given by $U = U_\text{PMNS}P$:
\begin{equation}
U = \begin{pmatrix}U_{e1}&U_{e2}&U_{e3}\\U_{\mu1}&U_{\mu2}&U_{\mu3}\\U_{\tau1}&U_{\tau2}&U_{\tau3}\end{pmatrix}
= \begin{pmatrix}
c_{12}c_{13}&s_{12}c_{13}&s_{13}e^{-i\delta}\\
-s_{12}c_{23}-c_{12}s_{13}s_{23}e^{i\delta}&c_{12}c_{23}-s_{12}s_{13}s_{23}e^{i\delta}&c_{13}s_{23}\\
s_{12}s_{23}-c_{12}s_{13}c_{23}e^{i\delta}&-c_{12}s_{23}-s_{12}s_{13}c_{23}e^{i\delta}&c_{13}c_{23}
\end{pmatrix},
\end{equation}
\begin{equation}
P = \diag(e^{i\alpha},e^{i\beta},1),
\end{equation}
here we used the notation of $c_{12}=\cos\theta_{12}$, $s_{12}=\sin\theta_{12}$ and so on. We identify $\theta_\odot$ and $\theta_\text{atm}$ as $\theta_{12}$ and $\theta_{23}$ respectively. And then we substitute the following neutrino oscillation data into $U_\text{PMNS}$:
\begin{align}
\varDelta m_\odot^2 &= \begin{matrix}7.9\pm0.3\\(7.1{-}8.9)\end{matrix}\times10^{-5},&
\sin^2\theta_\odot &= \begin{matrix}0.30^{+0.02}_{-0.25}\\(0.24{-}0.40)\end{matrix},\\
\abs{\varDelta m_\text{atm}^2} &= \begin{matrix}2.5^{+0.20}_{-0.25}\\(1.9{-}3.2)\end{matrix}\times10^{-3},&
\sin^2\theta_\text{atm} &= \begin{matrix}0.50^{+0.08}_{-0.07}\\(0.38{-}0.64)\end{matrix},
\end{align}
where the lower values are 95\% confidence intervals. Below, we concentrate the normal hierarchical mass spectrum of the light neutrinos, i.e. $m_1 \ll m_2 \ll m_3$. Thus we identify $\varDelta m_\odot^2$ and $\abs{\varDelta m_\text{atm}^2}$ as $m_2^2-m_1^2$ and $m_3^2-m_2^2 \simeq m_3^2-m_1^2$ respectively. And the worldwide reactor neutrino oscillation experiments gives the following observations:
\begin{equation}
\sin^2\theta_{13} <
\begin{cases}
0.027 (0.048) & \text{CHOOZ + atm. + LBL,}\\
0.033 (0.071) & \text{solar + KamLAND,}\\
0.020 (0.041) & \text{3--$\nu$ global data,}
\end{cases}
\label{ReactorAngle}
\end{equation}
here, the values in parentheses are 95\% upper confidence limits. Although we have five input parameters, $(m_1, \theta_{13}, \delta, \alpha, \beta)$, the lightest mass eigenvalue $m_1$ and the reactor neutrino angle $\theta_{13}$ are highly constrained. Hereafter we assume that $\theta_{13} = 0.02$. Then we have three free input parameters $(\delta, \alpha, \beta)$, which are all $\mathit{CP}$-violating phases.

In order to obtain the neutrino mass spectrum from these experimantal data, we require one more assumption. Then we use the up-down unification relation, i.e. the neutrino mass Dirac mass matrix \cite{MR}
\begin{equation}
M_\text{D} = \frac{m_t}{m_b}M_\ell = \frac{m_t}{m_b}\diag(m_e,m_\mu,m_\tau),
\end{equation}
here we use the basis where the charged lepton mass matrix $M_\ell$ is diagonal and real-valued. The GUT realization supports this assumption: the leptons belong to some large representation of the GUT group with the quarks. Thus, it seems to be natural that the lepton Dirac mass spectrum is proportional to the quark Dirac mass spectrum. Conbining the above considerations, we obtain
\begin{equation}
\begin{aligned}
M_R
&= \frac{m_t^2}{m_b^2}M_\ell{m_\nu}^{-1}M_\ell\\
&= \frac{m_t^2}{m_b^2}\begin{pmatrix}m_e&&\\&m_\mu&\\&&m_\tau\end{pmatrix}U_\text{PMNS}P^2\begin{pmatrix}m_1&&\\&m_2&\\&&m_3\end{pmatrix}\trans{U_\text{PMNS}}\begin{pmatrix}m_e&&\\&m_\mu&\\&&m_\tau\end{pmatrix}.
\end{aligned}
\end{equation}
This gives us the mass eigenvalues and phases $M_i = \abs{M_i}e^{i\phi_i/2}$ ($i = 1,2,3$). When we assume the Dirac and Majorana $\mathit{CP}$-phases $\delta$, $\alpha$ and $\beta$ to be $\delta = \alpha = \pi/2$ and $\beta = \pi$, we obtain the following values for $0.0001 \leq m_1 \leq 3$ [eV]:
\begin{equation}
\begin{aligned}
1.70\times10^8 &\lesssim M_1 \lesssim 1.55\times10^{10} \text{[GeV]},\\
3.52\times10^{10} &\lesssim M_2 \lesssim 1.96\times10^{12} \text{[GeV]},\\
2.03\times10^{13} &\lesssim M_3 \lesssim 5.25\times10^{14} \text{[GeV]}.
\end{aligned}
\label{HeavyNeutrinos}
\end{equation}
According to \cite{RHN1}, we have two choices:
\begin{enumerate}
\item Before starting the leptogenesis mechanism, there would exist no initial $N_1$-abundance. In case of the hierarchical neutrino mass spectrum $m_1 \ll m_2 \ll m_3$,  as we consider now, $M_1 \gtrsim 2.4\times10^9$ [GeV] is requred.
\item If the GUT interactions progress rapidly enough, $N_1$ is in thermal equilibrium at that temperature region. Then, $M_1 \gtrsim 4.9\times10^8$ [GeV] is compatible with the hierarchical neutrino mass spectrum.
\end{enumerate}
Summarizing these two constraints, in the following we consider $4.9\times10^8 \lesssim M_1 \lesssim 1.55\times10^{10}$ [GeV]. At last, we are ready to compute the generation of the net lepton and baryon numbers.

\section{Thermal leptogenesis}\label{Baryogenesis}
Our framework provides two sources of lepton asymmetry, i.e. the lighest heavy neutrino $N_1$ and the $\mathit{SU}(2)_L$ triplet scalar reperesentation $\Delta_L$. The produced $\mathit{CP}$-asymmetries are defined as \cite{triplet}
\begin{align}
\epsilon_{N_j} &\equiv \sum_i \frac{\varGamma(N_j \to \ell_i\overline{h})-\varGamma(N_j \to \overline{\ell}_ih)}{\varGamma(N_j \to \ell_i\overline{h})+\varGamma(N_j \to \overline{\ell}_ih)},\\
\epsilon_\Delta &\equiv 2\ \frac{\varGamma(\overline{\Delta_L} \to \ell_i\ell_j)-\varGamma(\Delta_L \to \overline{\ell}_i\overline{\ell}_j)}{\varGamma(\overline{\Delta_L} \to \ell_i\ell_j)+\varGamma(\Delta_L \to \overline{\ell}_i\overline{\ell}_j)},
\end{align}
here the symbol $h$ denotes the SM Higgs doublet $H$ or $H'$.

\begin{figure}[btp]
\begin{minipage}{0.4\textwidth}
\begin{center}
\includegraphics[width=\textwidth]{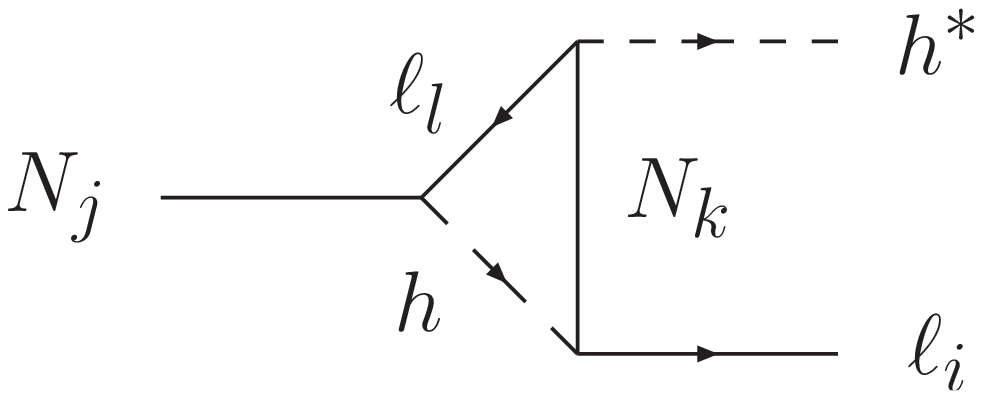}
\end{center}
\end{minipage}
\begin{minipage}{0.1\textwidth}
\hfil
\end{minipage}
\begin{minipage}{0.4\textwidth}
\begin{center}
\includegraphics[width=\textwidth]{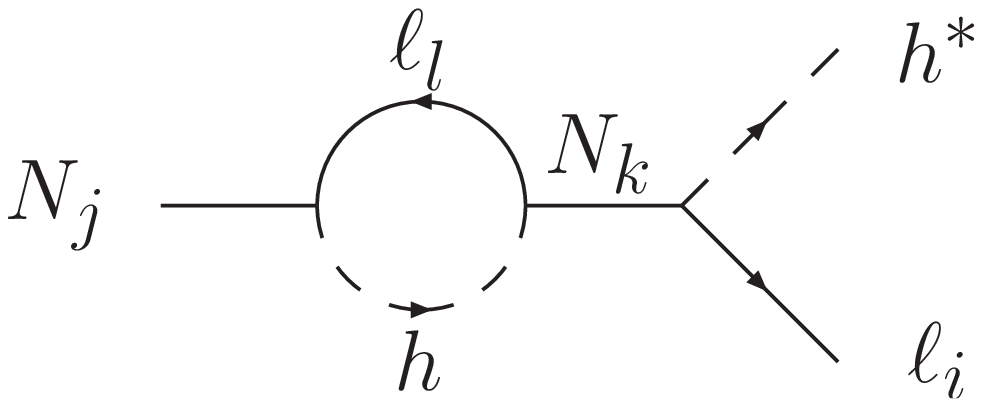}
\end{center}
\end{minipage}
\caption{Right-handed neutrinos decay contribution diagrams.}
\label{Ndecay1}
\end{figure}

First, let us consider the $N$-decaying contribution $\epsilon_{N_j}$~\cite{NDecay}. This can be computed as
\begin{align}
\epsilon_{N_j} &= \frac{1}{8\pi} \sum_k \frac{\Img[(Y^\dagger\tilde{Y})_{lk}(\tilde{Y}^\dagger Y)_{lk}]}{(Y^\dagger Y)_{jj}}f(x_k),&
f(x) &= \sqrt{x}\left\{1-(1+x)\ln\left(1+\frac{1}{x}\right)+\frac{1}{1-x}\right\},
\end{align}
where $x_k \equiv M_k^2/M_j^2$. We can replace the Yukawa couplings by the mass matrix:
\begin{equation}
\epsilon_{N_j} = \frac{1}{8\pi\kappa_+^2({M_\text{D}}^\dagger{M_\text{D}})_{jj}} \sum_{k \neq j} \Img[({M_\text{D}}^\dagger{M_\text{D}})_{jk}]^2f(x_k).
\end{equation}
As remarked in the previous section, the heavy neutrino mass spectrum is hierarchical. This suggests that while the heavier $N_2$ and $N_3$ are decaying, the lightest $N_1$ is still in equilibrium. In other words, the lepton number asymmetry generated by $N_2$ and $N_3$-decaying processes should be erased by the lepton number violating scatterings according to the presence of $N_1$. Then we find that only the $N_1$-decaying processes are dominant in the asymmetry $\epsilon_{N_j}$:
\begin{equation}
\sum_{N_j} \epsilon_{N_j} \simeq \epsilon_{N_1} = -\frac{3}{16\pi\kappa_+^2({M_\text{D}}^\dagger{M_\text{D}})_{11}} \sum_{k=2,3} \Img[({M_\text{D}}^\dagger{M_\text{D}})^2_{1k}]\frac{M_1}{M_k}.
\label{CP1}
\end{equation}

\begin{figure}[btp]
\includegraphics[width=0.4\textwidth]{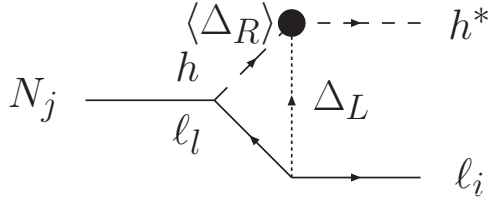}
\caption{Additional one-loop contribution to right-handed neutrinos decay.}
\label{Ndecay2}
\end{figure}

Next, we consider the $N_j$-decay including the virtual $\Delta_L$ as tipified by Fig.~\ref{Ndecay2}:
\begin{align}
\epsilon_{N_j}^\Delta &= -\frac{1}{2\pi} \sum_{k,l} \frac{\Img[Y^\ast_{lj}\tilde{Y}^\ast_{kj}{Y_\Delta}_{lk}\beta v_R]}{M_j(Y^\dagger Y)_{jj}}g(x_j),&
g(x_j) &= 1-\frac{M_{\Delta_L}^2}{M_j^2} \ln\left(1+\frac{M_j^2}{M_{\Delta_L}^2}\right).
\end{align}
Substituting the couplings for mass matrices, we find
\begin{equation}
\epsilon_{N_j}^\Delta = -\frac{1}{2\pi v_L M_j({M_\text{D}}^\dagger{M_\text{D}})_{jj}} \sum_{k,l} \Img[({M_\text{D}})^\ast_{lj}({M_\text{D}})^\ast_{kj}(m_\nu^\text{\Rnum{2}})_{lk}\beta v_R]g(x_j).
\label{CP2}
\end{equation}

\begin{figure}[tbp]
\includegraphics[width=0.4\textwidth]{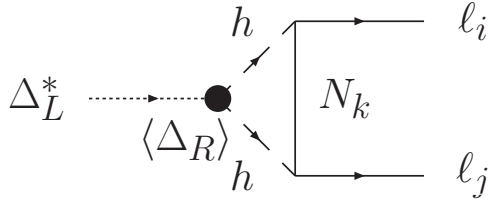}
\caption{Left-handed triplet scalar decay contribution diagrams.}
\label{Deltadecay}
\end{figure}

Our last task is computing the $\Delta_L$-decaying contribution such as Fig.~\ref{Deltadecay}. This is given by
\begin{equation}
\epsilon_\Delta = \frac{1}{8\pi}\sum_k M_k\frac{\sum_{i,j} \Img[Y^\ast_{ik}\tilde{Y}^\ast_{jk}\beta^\ast v_R{Y_\Delta}_{ij}]}{\sum_{i,j} \abs{{Y_\Delta}_{ij}}^2M_{\Delta_L}^2+\sum_{a,b}\abs{\beta_{ab}}^2v_R^2} \ln\left(1+\frac{M_{\Delta_L}^2}{M_k^2}\right).
\end{equation}

\subsection{SM-type scenario}
At first we consider the parametrization showed in Fig.~\ref{RGE1}, where $M_{\Delta_L} = 6\times10^9$ [GeV]. This parametrization requires a huge mass hierarchy in doublet Higgs sector: $m_H \ll m_{H'}$. If we employ the smallest value of $M_1$ from Eq.~\eqref{HeavyNeutrinos}, we find that lept on asymmetry is effectively generated in the temperature range of
\begin{equation}
100 \ \text{[GeV]} \sim T_\text{EW} < T_\text{sph}^\text{SM} < M_1. \label{SMSph}
\end{equation}
According to Fig.~\ref{RGE1}, the surviving gauge symmetry in this region is the SM one. Then in addition to the Sphaleron processes, all the interactions in the ordinary SM are in thermal equilibrium. These in-equilibrium interactions interrelate each chemical potentials of the particles. Furthermore since the universe has to be neutral, then the conserved charges, i.e. the third component of the left-handed isospin $I_L^3$ and the hypercharge $Y$ are to be zero respectively. $I_L^3$ and $Y$ are given by
\begin{align}
I_L^3 &= \frac{g_L^2T^3}{6}\left\{\frac{1}{2}\sum_i\left(\sum_\text{color}(\mu_{u_L^i}-\mu_{d_L^i})+(\mu_{\nu_L^i}-\mu_{e_L^i})\right)+(\mu_{h^+}-\mu_{h^0})+4\mu_{W^+}\right\},\label{SM-I}\\
Y &= \frac{g_Y^2T^3}{6}\left[\sum_i\left\{\sum_\text{color}\left(\frac{1}{3}(\mu_{u_L^i}+\mu_{d_L^i})+\frac{4}{3}\mu_{u_R^i}-\frac{2}{3}\mu_{d_R^i}\right)-(\mu_{\nu_L^i}+\mu_{e_L^i})-2\mu_{e_R^i}\right\}+2(\mu_{h^+}+\mu_{h^0})\right].\label{SM-Y}
\end{align}
In the temerature region \eqref{SMSph}, since the $\mathit{SU}(2)_L$ gauge interactions of up-type quarks are in thermal equilibrium, all $u_L^i$ are mixed enough. Therefore we can not distinguish the chemical potentials of up-type quarks. Then we have $\mu_{u_L} \equiv \mu_{u_L^i}$ ($i=1,2,3$). Similarly we can ignore the generation index of down-type quarks, then we obtain $\mu_{d_L} \equiv \mu_{d_L^i}$ ($i=1,2,3$). The in-equilibrium Yukawa interactions lead $\mu_{u_R} \equiv \mu_{u_R^i}$ and $\mu_{d_R} \equiv \mu_{d_R^i}$ ($i=1,2,3$). The same can be said for the lepton sector, we obtain $\mu_\nu \equiv \mu_{\nu_L^i}$, $\mu_{e_L} \equiv \mu_{e_L^i}$ and $\mu_{e_R} \equiv \mu_{e_R^i}$. Consequently we find $\mu_B = \mu_{W^0} = \mu_{W^+} = 0$. Since $\mathit{SU}(2)_L\!\times\!\mathit{U}(1)_Y$ gauge interactions are in thermal equilibrium, this is exactly what we expected. And we obtain the well-known formula
\begin{equation}
B = \frac{28}{79}(B-L) = -\frac{28}{51}L.
\label{BL:SM}
\end{equation}

Since $M_1 < M_{\Delta}$, using Eq.~\eqref{vL}, we can werite Eq.~\eqref{CP2}:
\begin{equation}
\sum_{N_j} \epsilon_{N_j}^\Delta \simeq \epsilon_{N_1}^\Delta = -\frac{3M\eta_\text{P}}{8\pi \kappa_+^2M_1({M_\text{D}}^\dagger{M_\text{D}})_{11}} \sum_{k,l} \Img[({M_\text{D}})^\dagger_{1l}(m_\nu^\text{\Rnum{2}})_{lk}({M_\text{D}})^\ast_{k1}],
\end{equation}
Here we consider that the dominant $CP$-asymmetry comes from the $N_1$-decay, i.e. $\epsilon \sim \epsilon_{N_1}$. Then the net lepton number is produced by ordinary $N_1$-decaying process and the lepton number violating scattering. This situation has been extensively studied \cite{leptogenesis2}. The Boltzmann equation (BE) for $N_1$-abundance is written as
\begin{align}
\frac{d\tilde{Y}_{N_1}}{dz}
&= -z\frac{\expect{\Gamma_{N_1}}_z}{H(z=1)}\left(\tilde{Y}_{N_1}-\tilde{Y}_{N_1}^\text{eq}\right),&
\end{align}
where $\tilde{Y}_{N_1} \equiv g_\ast n_{N_1}/s$ and $\tilde{Y}_{N_1}^\text{eq} \equiv g_\ast n_{N_1}^\text{eq}/s$. Here we use a dimensionless evolution parameter $z=M_1/T$. And $\expect{\Gamma_{N_1}}(z)$, $H(z)$ and $Y_i^\text{eq}(z)$ are the Maxwell-Boltzmann averaged total decay rate of $N_1$, the Hubble parameter and the equilibrium yield variable of a particle spieces $i$ respectively. The net lepton number density $n_L = n_\ell-n_{\overline{\ell}}$ evolves as
\begin{equation}
\begin{aligned}
\frac{d\tilde{Y}_L}{dz}
&= z\frac{\expect{\Gamma_{N_1}}_z}{H(z=1)}\left(\tilde{Y}_{N_1}-\tilde{Y}_{N_1}^\text{eq}\right)\\
&\quad -\frac{12\zeta(3)}{\pi^2}z\frac{\expect{\Gamma_{N_1}}_z}{H(z=1)}\left(\tilde{2Y}_L+\tilde{Y}_{h1}\right)\left(\frac{\pi^2}{2\zeta(3)}\frac{\expect{\Gamma_{N_1}}\tilde{Y}_{N_1}^\text{eq}}{g_\ast T^3}+\frac{45}{2\pi^2}\frac{2\zeta(3)}{\pi^2 g_{\ast S}}\expect{\sigma'\abs{v}}\right).
\end{aligned}
\end{equation}
where $\tilde{Y}_L$ is defined by $g_\ast/\epsilon \cdot n_L/s$. And $\expect{\sigma'\abs{v}}$ denotes the thermal averaged cross section of the $\varDelta L =2$ scatterings in the thermal bath. Furthermore since $\overline{H}$ always appears with a lepton $\ell$, the BE for the net Higgs number density $n_{h1} \equiv n_H-n_{\overline{H}}$ is identical to that for $n_L$.

\begin{figure}[htbp]
\begin{minipage}{0.49\textwidth}
\begin{center}
\includegraphics[width=\textwidth]{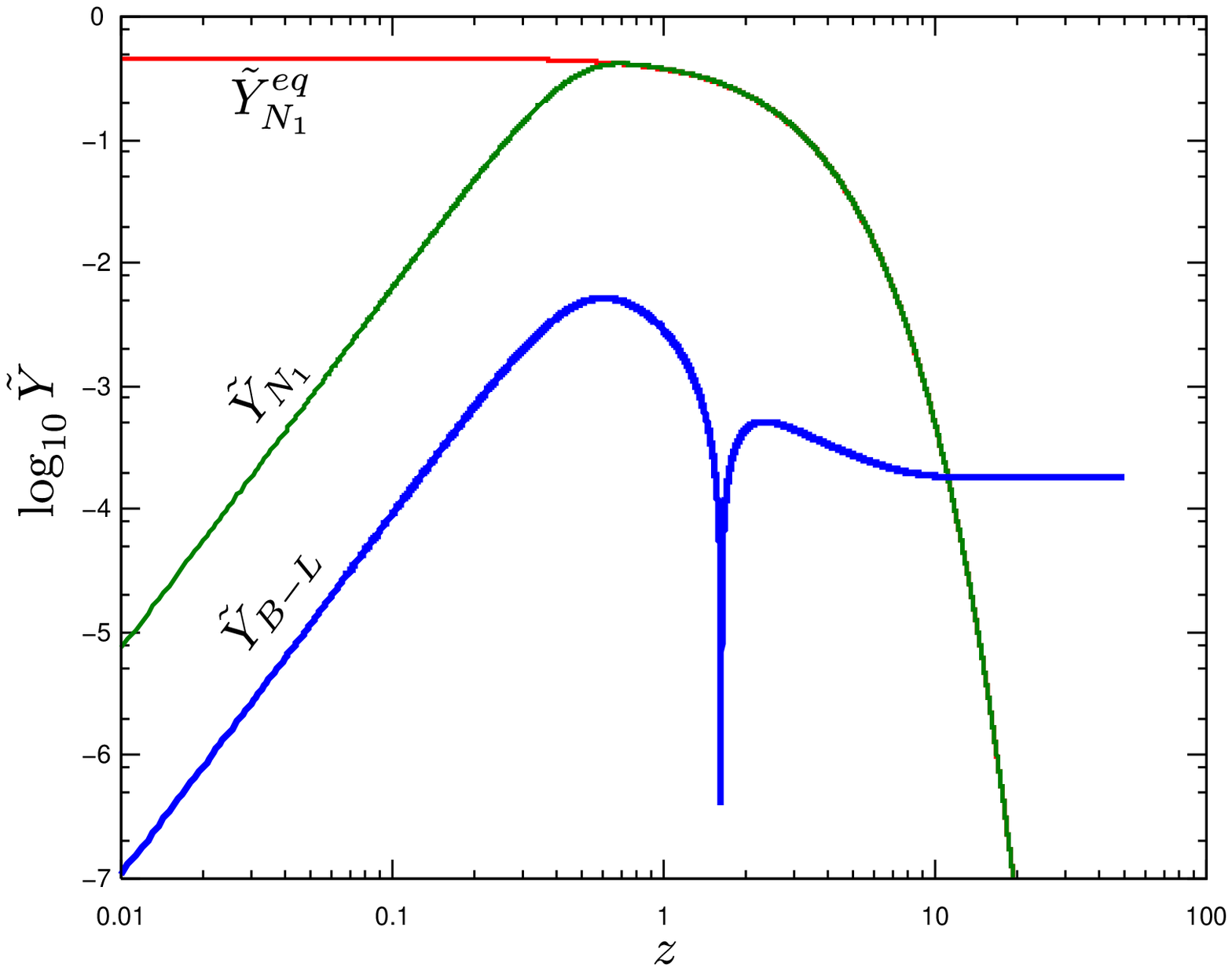}
\caption{\def\baselinestretch{.7}\normalsize Evolution of $Y_{N_1}(z)$, $Y_{N_1}^\text{eq}(z)$ and $Y_{B{-}L}(z)$ without initial $N_1$-abundance.}
\label{BE-SM1}
\end{center}
\end{minipage}
\begin{minipage}{0.49\textwidth}
\begin{center}
\includegraphics[width=\textwidth]{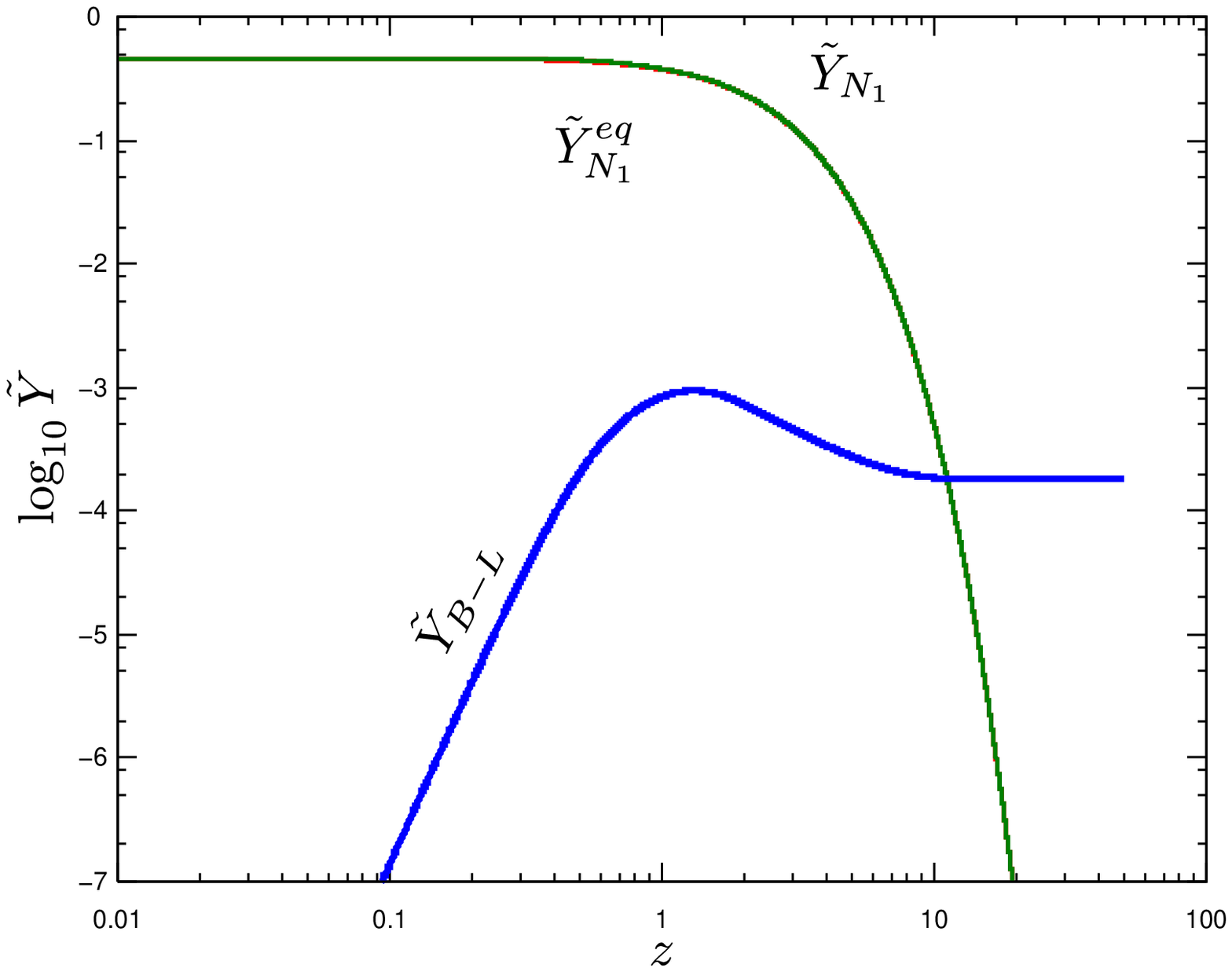}
\caption{\def\baselinestretch{.7}\normalsize Evolution of $Y_{N_1}(z)$, $Y_{N_1}^\text{eq}(z)$ and $Y_{B{-}L}(z)$ with initial $N_1$ at equilibrium.}
\label{BE-SM2}
\end{center}
\end{minipage}
\end{figure}

Now we consider the lepton symmetric universe with no doublet: $n_L(z \to 0) = n_{h1}(z \to 0) = 0$. For example, in order to obtain successful baryogenesis we take $m_1 \sim 0.0020197$ [eV], $m_2 = 0.0086557$ [eV], $m_3 = 0.049245$ [eV]. Then this choice gives
\begin{align}
M_1 &= 3.6892\times10^8 \text{[GeV]},&
M_2 &= 1.0103\times10^{10} \text{[GeV]},&
M_3 &= 1.2725\times10^{14} \text{[GeV]}.
\label{MR}
\end{align}
and $\abs{\epsilon_{N_1}} \simeq  1.2135\times10^{-8}$. Then we find $\eta_B = 5.8556\times10^{-10}$ (see Fig.~\ref{BE-SM1}). In Fig.~\ref{BE-SM1} the red and the green curves represent for $Y_{N_1}^\text{eq}(z)$ and $Y_{N_1}(z)$ respectively. The abundance of $B{-}L$ changes it own sign during leptogenesis, then $\abs{Y_{B{-}L}}(z)$ is plotted with a blue curve. This value of $M_1$ is lower than the lower bound in case of no initial $N_1$-abundance.

Next, we assume that $N_1$ have already been in equilibrium at $z \to 0$. The mass spectrum \eqref{MR} gives $\eta_B = 5.8561\times10^{-10}$. $M_1$ in Eq.~\eqref{MR} is near the lowest value in the case of the initially thermal $N_1$-abundance. We show the time-evolution in Fig.~\ref{BE-SM2}.

\subsection{2HDM-type scenario}
Now let us expand our discussion one step further. Then we consider the parametrization showed in Fig.~\ref{RGE2}, where $M_{\Delta_L} = 1.0\times10^{10}$ [GeV]. Differently from the previous case, we need no fine-tuning in mass spectrum of doublets. Since $M_1 < M_{\Delta_L}$, the the lepton asymmetry is produced in the range of
\begin{equation}
100 \ \text{[GeV]} \sim T_\text{EW} < T_\text{sph}^\text{2HDM} < M_1. \label{2HDMSph}
\end{equation}
Since in addition to the SM the extra Dirac--Yukawa interactions are also in thermal equilibrium, the baryon conversion ratio is modified a little. The second Higgs doublet $H'$ behaves like a copy of $\overline{H}$ in terms of chemical potential. Then ignoring the family index same as before, we can replace Eq.~\eqref{SM-I} and \eqref{SM-Y} by
\begin{align}
I_L^3 &= \frac{g_L^2T^3}{6}\left\{\frac{1}{2}\sum_i\left(\sum_\text{color}(\mu_{u_L^i}-\mu_{d_L^i})+(\mu_{\nu_L^i}-\mu_{e_L^i})\right)+2(\mu_{h^+}-\mu_{h^0})+4\mu_{W^+}\right\},\label{2HDM-I}\\
Y &= \frac{g_Y^2T^3}{6}\left[\sum_i\left\{\sum_\text{color}\left(\frac{1}{3}(\mu_{u_L^i}+\mu_{d_L^i})+\frac{4}{3}\mu_{u_R^i}-\frac{2}{3}\mu_{d_R^i}\right)-(\mu_{\nu_L^i}+\mu_{e_L^i})-2\mu_{e_R^i}\right\}+4(\mu_{h^+}+\mu_{h^0})\right].\label{2HDM-Y}
\end{align}
resspectively. Using these equations, we obtain
\begin{equation}
B = \frac{8}{23}(B-L) = -\frac{8}{15}L.
\label{BL:2HDM}
\end{equation}
Since the doublet Higgs bosons do not have $B{-}L$ charges, we find that the baryon convertion factor $C_\text{sph}^\text{2HDM} = 8/23 \simeq 0.348$ is about as same as $C_\text{sph}^\text{SM} = 28/79 \simeq 0.354$.

\begin{figure}[htbp]
\begin{minipage}{0.49\textwidth}
\begin{center}
\includegraphics[width=\textwidth]{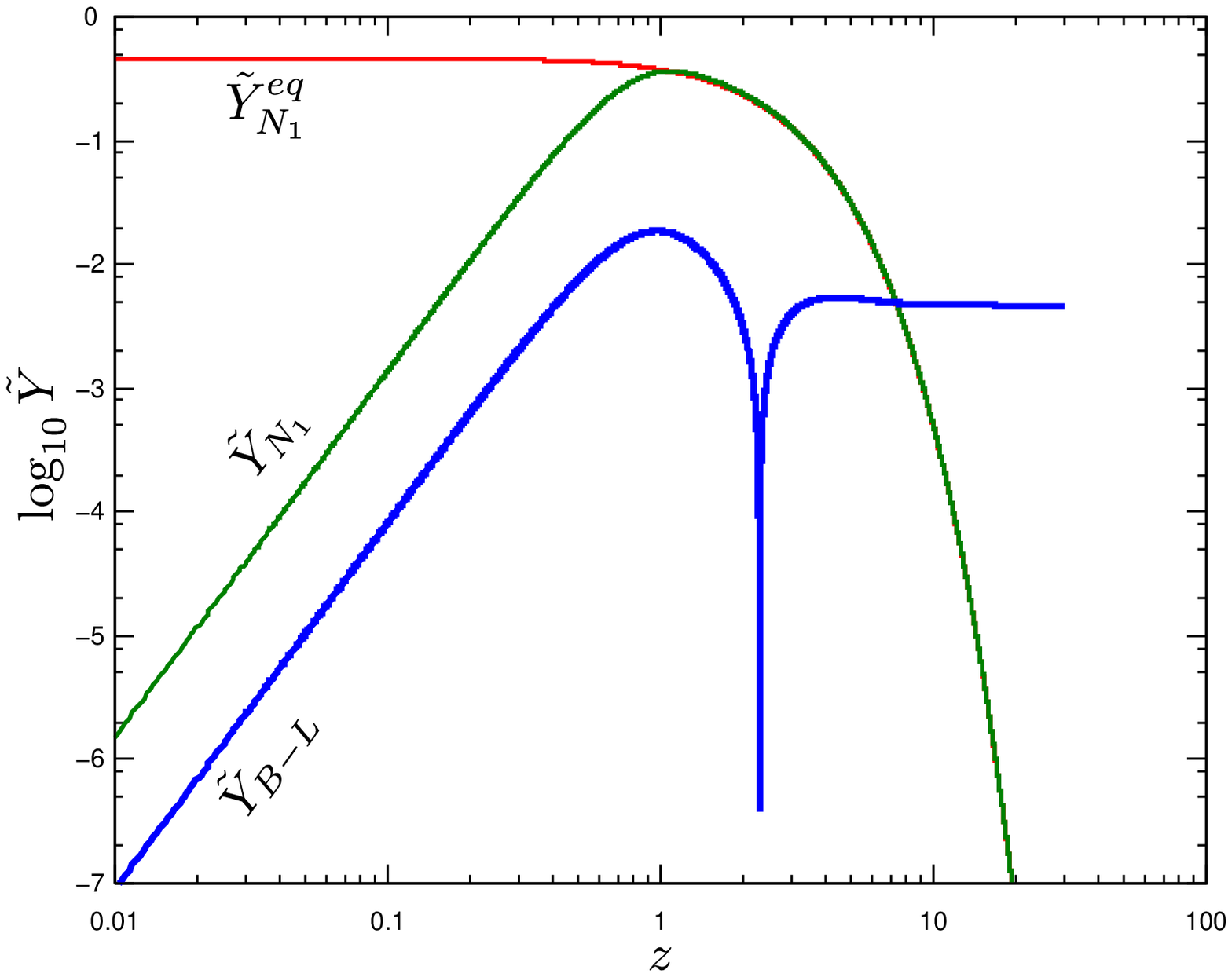}
\caption{\def\baselinestretch{.7}\normalsize Evolution of $Y_{N_1}(z)$, $Y_{N_1}^\text{eq}(z)$ and $Y_{B{-}L}(z)$ without initial $N_1$-abundance.}
\label{BE-2HDM1}
\end{center}
\end{minipage}
\begin{minipage}{0.49\textwidth}
\begin{center}
\includegraphics[width=\textwidth]{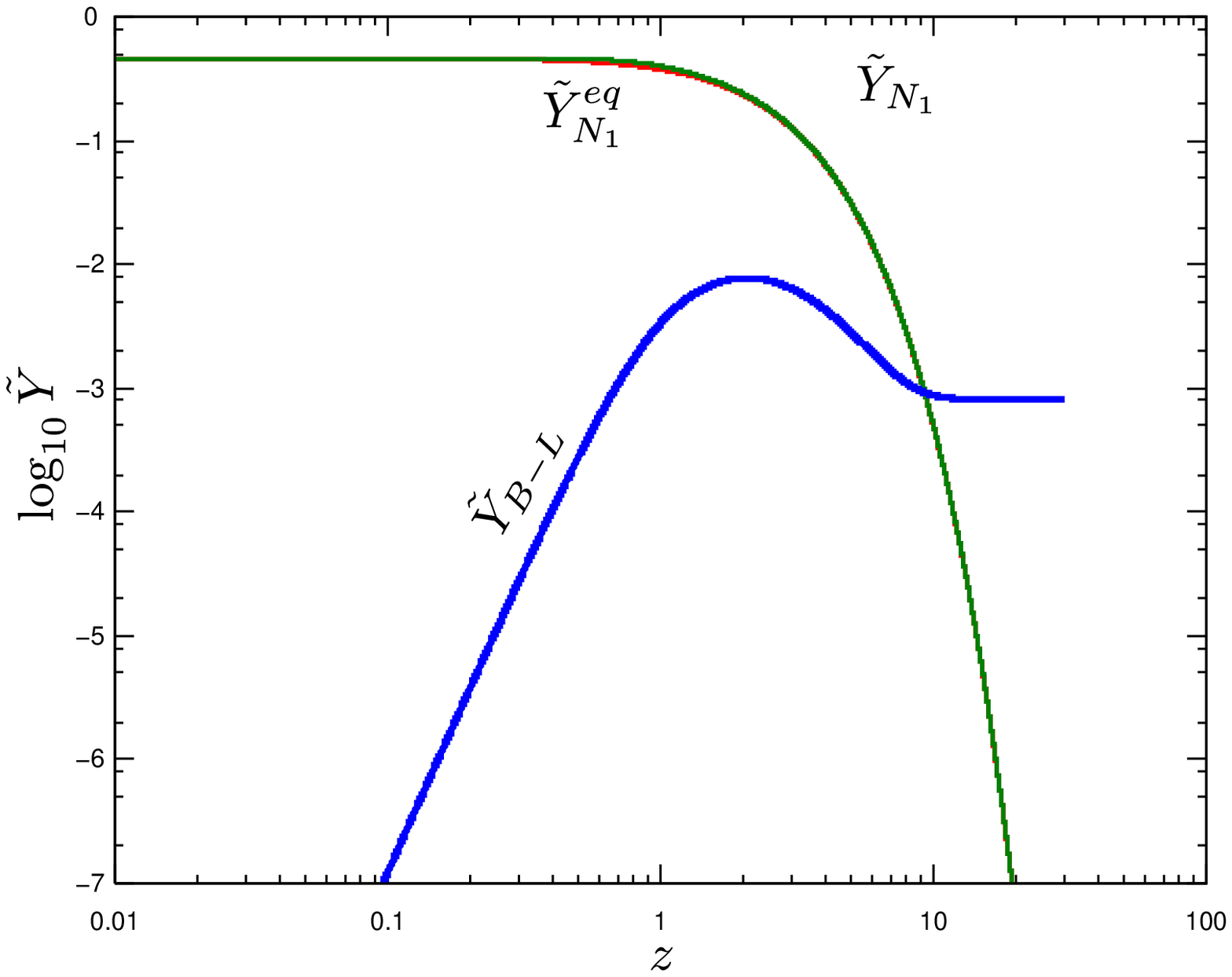}
\caption{\def\baselinestretch{.7}\normalsize Evolution of $Y_{N_1}(z)$, $Y_{N_1}^\text{eq}(z)$ and $Y_{B{-}L}(z)$ with initial $N_1$ at equilibrium.}
\label{BE-2HDM2}
\end{center}
\end{minipage}
\end{figure}

This parametrization leads that the expected $CP$-asymmetry is double the previous SM case \eqref{CP1}. Therefore this fact relaxes the constraint on $m_1$ for successful leptogenesis. Since $H'$ behaves as a copy of $\overline{H}$, we can introduce $Y_\Phi \equiv Y_{h1}+Y_{h2} = 2Y_{h1}$, where $Y_{h2}$ denotes the net second Higgs number yield variable: $Y_{h2} \equiv (n_{H'^\ast}-n_{H'})/s$. For example, let us consider $m_1 = 10^{-4}$ [eV]. We show the results in Fig.~\ref{BE-2HDM1} and \ref{BE-2HDM2}. We obtain $\eta_B = 9.5634\times10^{-9}$ and $1.6560\times10^{-9}$ respectively.

\subsection{Leptogenesis through $\Delta_L$-decaying process}
In the following, let us consider the case where the dominant $\mathit{CP}$-asymmetry comes from $\Delta_L$-decay, i.e. $\epsilon \sim \epsilon_\Delta$. And we consider the type-\Rnum{1} seesaw mechanism \eqref{type1}. The case of the type-\Rnum{2} has been studied in \cite{Antusch}. We can imagine this case if $M_{\Delta_L} \lesssim M_1$. As we showed before, Fig.~\ref{GUTs} shows that the mass of $\Delta_L$ does not affect whether the gauge unification would come true. Let us assume the SM with a sufficiently light triplet (SM+$\Delta$), $M_{\Delta_L} \sim \mathcal{O}(M_W)$. Fig.~\ref{GUTs} allows $\Delta_L$ to live in the electroweak scale, while the hierarchical relation $m_H \ll m_{H'}$ requires a fine-tuning. In the region of $T > M_{\Delta_L}$, the following interactions are in thermal equilibrium \cite{DeltaL}:
\renewcommand{\labelenumi}{\theenumi}
\renewcommand{\theenumi}{(\alph{enumi})}
\begin{enumerate}
\item Gauge interactions of triplets.
\begin{subequations}
\begin{align}
\Delta^-_L+\Delta^0_L &\leftrightarrow W^-_L,&
\mu_{\Delta^+} &= \mu_{\Delta^0}+\mu_{W^+},\\
\Delta^{--}_L+\Delta^+_L &\leftrightarrow W^-_L,&
\mu_{\Delta^{++}} &= \mu_{\Delta^+}+\mu_{W^+}, 
\end{align}
\label{TripletGaugeChem}
\end{subequations}
\label{TripletGauge}
\item Majorana--Yukawa couplings of leptons.
\begin{subequations}
\begin{align}
\Delta^0_L &\leftrightarrow \overline{\nu}^i+\overline{\nu}^j,&
\mu_{\Delta^0} &= -2\mu_\nu,\\
\Delta^+_L &\leftrightarrow \overline{\nu}^i+\overline{e_L}^j,&
\mu_{\Delta^+} &= -\mu_\nu-\mu_{e_L},\\ 
\Delta^{++}_L &\leftrightarrow \overline{e_L}^i+\overline{e_L}^j,&
\mu_{\Delta^{++}} &= -2\mu_{e_L}, 
\end{align}
\label{TripletMajoranaChem}
\end{subequations}
\label{TripletMajorana}
\item Cubic interactions between two doublets and a triplet.
\begin{subequations}
\begin{align}
\Delta^0_L &\leftrightarrow h^0+h^0,&
\mu_{\Delta^0} &= 2\mu_{h^0},\\
\Delta^+_L &\leftrightarrow h^0+h^+,&
\mu_{\Delta^+} &= \mu_{h^0}+\mu_{h^+},\\ 
\Delta^{++}_L &\leftrightarrow h^++h^+,&
\mu_{\Delta^{++}} &= 2\mu_{h^+}, 
\end{align}
\label{TripletCubic1Chem}
\end{subequations}
\label{TripletCubic1}
\end{enumerate}
Note that the above interactions violate the lepton number explicitly. The interaction \ref{TripletCubic1} prevents a dangerous Majoron. This is motivated by astrophysical reason and $Z^0$ total width data at LEP. And the neutralness under the $\mathit{SU}(2)_L$ and $\mathit{U}(1)_Y$ gauge symmetries are guaranteed by
\begin{align}
0 = I_L^3 &\propto \frac{1}{2}\sum_i\left(\sum_\text{color}(\mu_{u_L^i}-\mu_{d_L^i})+(\mu_{\nu_L^i}-\mu_{e_L^i})\right)+(\mu_{h^+}-\mu_{h^0})-2\mu_{\Delta^0}+2\mu_{\Delta^{++}}+4\mu_{W^+},\label{SM+triplet-I}\\
0 = Y &\propto \sum_i\left\{\sum_\text{color}\left(\frac{1}{3}(\mu_{u_L^i}+\mu_{d_L^i})+\frac{4}{3}\mu_{u_R^i}-\frac{2}{3}\mu_{d_R^i}\right)-(\mu_{\nu_L^i}+\mu_{e_L^i})-2\mu_{e_R^i}\right\}+2(\mu_{h^+}+\mu_{h^0}) \notag\\
&\quad +4(\mu_{\Delta^0}+\mu_{\Delta^+}+\mu_{\Delta^{++}}).\label{SM+triplet-Y}
\end{align}
As a result, we obtain
\begin{equation}
B = \frac{11}{20}(B-L) = -\frac{11}{9}L.
\end{equation}
Unlike the previous values, this convertion factor $C_\text{sph}^\text{SM+$\Delta$} = 11/20 = 0.55$ is relatively large. The nonzero $B{-}L$ charge of $\Delta_L$ causes this result.

If the second doublet $H'$ also has an electroweak scale mass, the effective theory becomes the 2HDM with a light triplet (2HDM+$\Delta$). This configuration requires no fine-tuning $m_H$ and $m_{H'}$. In this case we obtain the followiing relation between $B$ and $L$ in the range of $T > M_{\Delta_L}$:
\begin{equation}
B = \frac{16}{43}(B-L) = -\frac{16}{27}L.
\end{equation}
This relation can be obtained by using the following equations in addition to \ref{TripletGauge}, \ref{TripletMajorana} and \ref{TripletCubic1}:
\begin{enumerate}\setcounter{enumi}{3}
\item Cubic interactions between two second doublets and a triplet.
\begin{subequations}
\begin{align}
\Delta^0_L &\leftrightarrow h'^0+h'^0,&
\mu_{\Delta^0} &= 2\mu_{h'^0},\\
\Delta^+_L &\leftrightarrow h'^0+h'^+,&
\mu_{\Delta^+} &= \mu_{h'^0}+\mu_{h'^+},\\ 
\Delta^{++}_L &\leftrightarrow h'^++h'^+,&
\mu_{\Delta^{++}} &= 2\mu_{h'^+}, 
\end{align}
\label{TripletCubic2Chem}
\end{subequations}
\label{TripletCubic2}
\end{enumerate}
And the global neutralness conditions are given by
\begin{align}
0 = I_L^3 &\propto \frac{1}{2}\sum_i\left(\sum_\text{color}(\mu_{u_L^i}-\mu_{d_L^i})+(\mu_{\nu_L^i}-\mu_{e_L^i})\right)+2(\mu_{h^+}-\mu_{h^0})-2\mu_{\Delta^0}+2\mu_{\Delta^{++}}+4\mu_{W^+},\label{2HDM+triplet-I}\\
0 = Y &\propto \sum_i\left\{\sum_\text{color}\left(\frac{1}{3}(\mu_{u_L^i}+\mu_{d_L^i})+\frac{4}{3}\mu_{u_R^i}-\frac{2}{3}\mu_{d_R^i}\right)-(\mu_{\nu_L^i}+\mu_{e_L^i})-2\mu_{e_R^i}\right\}+4(\mu_{h^+}+\mu_{h^0}) \notag\\
&\quad +4(\mu_{\Delta^0}+\mu_{\Delta^+}+\mu_{\Delta^{++}}).\label{2HDM+triplet-Y}
\end{align}
We find that the baryon convertion rate $C_\text{sph}^\text{2HDM+$\Delta$}$ is also about as same as $C_\text{sph}^\text{SM} $.

As is leptogenesis through $N_1$-decay, $\Delta_L$ needs to decouple from thermal bath at high temperature. This decoupling condition is given by $\expect{\Gamma_{\Delta_L}}_{T=M_{\Delta_L}} \lesssim H(T=M_{\Delta_L})$. Before solving the BE's, let us investigate this condition by order estimation. The tree level total decay rate of the triplet scalar $\Delta_L$ is given by
\begin{equation}
\expect{\Gamma_{\Delta_L}}_{T=M_{\Delta_L}} = \frac{K_1(T=M_{\Delta_L})}{K_2(T=M_{\Delta_L})}\frac{M_{\Delta_L}}{8\pi}\left(\sum_{i,j}\abs{{Y_\Delta}_{ij}}^2+\frac{\sum_{a,b}\abs{\beta_{ab}}^2v_R^2}{M_{\Delta_L}^2}\right),
\end{equation}
and the Hubble parameter at temperature $T$ can be written as
\begin{equation}
H(T) = \sqrt{\frac{4\pi^3g_\ast}{45}}\frac{T^2}{M_\text{P}},
\end{equation}
where $M_\text{P} = 1.22\!\times\!10^{19}$ [GeV]. If $\sum_{i,j}\abs{{Y_\Delta}_{ij}}^2 \gtrsim \sum_{a,b}\abs{\beta_{ab}}^2v_R^2M_{\Delta_L}^2$ and $\beta_{ab} = \mathcal{O}(1)$, the decoupling condition and our RG-analysis suggests that $M_{\Delta_L}$ is larger than at least $6.60\!\times\!10^{19}$ GeV (the SM+$\Delta$) or $1.24\!\times\!10^{22}$ GeV (the 2HDM+$\Delta$). This clearly conflicts with $M_{\Delta_L} \lesssim M_1$. Consequently we consider that the term proportional to $v_R^2/M_{\Delta_L}^2$ gives a dominant contribution, then we have
\begin{equation}
\expect{\Gamma_{\Delta_L}}_{T=M_{\Delta_L}} \simeq \frac{K_1(T=M_{\Delta_L})}{K_2(T=M_{\Delta_L})}\frac{v_R^2}{8\pi M_{\Delta_L}}.
\end{equation}
This means that
\begin{displaymath}
\sum_{i,j}\abs{{Y_\Delta}_{ij}}^2 = \frac{1}{v_R^2}\sum_{i,j}\abs{{M_R}_{ij}}^2 \sim \frac{M_3^2}{v_R^2} \ll \frac{v_R^2}{M_{\Delta_L}^2},
\end{displaymath}
then we obtain
\begin{equation}
M_{\Delta_L} \lesssim v_R^2/M_3.
\label{MDL0}
\end{equation}
Our RG-analysis suggests that $100 \ \text{[GeV]} < M_{\Delta_L} < \mathcal{O}(10^{5\text{--}10}) \ \text{[GeV]}$ for the SM+$\Delta$ and $100 \ \text{[GeV]} < M_{\Delta_L} < \mathcal{O}(10^{12\text{--}14}) \ \text{[GeV]}$ for the 2HDM+$\Delta$. Here lets us estimate the consistency of the above relation and our model. In addition to the decoupling of $\Delta_L$, at temperature $T = M_{\Delta_L}$ $W$-boson also has to be out-of-equilibrium. This gives another decoupling condition $\expect{\Gamma_W}_{T=M_{\Delta_L}} \lesssim H(T=M_{\Delta_L})$. According to \cite{Narendra}, we have $M_{\Delta_L} \gtrsim 4.8\!\times\!10^{10}$ [GeV]. Then it can be summarized as follows:
\begin{subequations}
\begin{align}
M_{\Delta_L} &= \mathcal{O}(10^{10}) \ \text{[GeV]}&
&\text{for the SM+$\Delta$},\label{MDL1}\\
M_{\Delta_L} &= \mathcal{O}(10^{10\text{--}14}) \ \text{[GeV]}&
&\text{for the 2HDM+$\Delta$}.\label{MDL2}
\end{align}
\end{subequations}
Eq.~\eqref{MDL1} is a apparently strong constraint. However, when we feed this and $v_R = \mathcal{O}(10^9)$ [GeV] to Eq.~\eqref{MDL0}, we obtain $M_3 \lesssim \mathcal{O}(10^8)$ [GeV]. This upper bound of $M_3$ conflicts with the hierarchical spectrum \eqref{HeavyNeutrinos}. Therefore we can conclude that leptogenesis through $\Delta_L$-decay process is impossible in the case of the hierarchical neutrino mass spectrum in the SM+$\Delta$. While in the 2HDM+$\Delta$ Eq.~\eqref{MDL2} is consistent with the hierarchical spectrum \eqref{HeavyNeutrinos}. In this case, we can obtain the suncessful value of $\eta_B$ if $m_1 \ll 10^{-4}$ [eV].

\section{Conclusion and future issues}
We found that the allowed region in the $(M_{Z_R}, M_{\Delta_L})$ parameter space is highly constrained in our model. This time, we focused on the nonsupersymmetric versions, and studied two specific cases, $M_{\Delta_L} = 200$ [GeV] and $M_{\Delta_L} = M_{Z_R} \approx M_{\Delta_R}$. As a result, we found that thermal $N_1$-leptogenesis scenario is successful in both the SM and the 2HDM.  In addition, we found that thermal $\Delta_L$-leptogenesis scenario accord with the hierarchical neutrino spectrum only in the 2HDM+$\Delta$, While thermal leptogenesis through $\Delta_L$-decay is incompatible with the hierarchical neutrinos in the SM. Here, it should be kept in mind that though we can build the more constrained models by the numerical analysis of RGE's we have no guide principle for $\mathcal{O}(M_{\Delta_L})$. We achieved our original target of building constrained LR models.

Our next step is to investigate under the framework of the finite-temperature field theory. In the literature, thermal leptogenesis scenarios are computed by the usual zero-temperature field theory. By including the thermal corrections we can discuss truely ``thermal'' leptogenesis processes. In \cite{TFT} the thermal $N_1$-leptogenesis in the SM and the MSSM is analysed using the finite-temperature field theory. We are now recomtuting our models using the Keldysh (real time) formalism \cite{TFT,RTF}. This approach can make more faithful predictions about not only leptogenesis scenatios but also the doublet Higgs sector. The above approach can be applied to the PS models and the $\mathit{SO}(10)$ GUT models. While the other approach is to introduce the supersymmetry. Although the supersymmetric extension of LR models (LRSUSY) \cite{LRSUSY} are compatible with $\mathit{R}$-parity conservation, unfortunately the LRSUSY models have many parameters, especially the soft breaking couplings make it difficult to predictive discussions. Furthermore, supersymmetry provides a new candidate of leptogenesis, i.e. Affleck--Dine leptogenesis. We are also interested in the competition between thermal leptogenesis \cite{leptogenesis3} and Affleck--Dine leptogenesis \cite{AD} in the LRSUSY models. Even if one could generate enough baryon asymmetry, the gravitino and reheating problems \cite{reheating} remain to be solved. We think our study as the first step in developping predictive models. These bottom-up approaches could give suggestive information on the Majorana couplings and the scalar four-point couplings.

\begin{acknowledgments}
I would like to thank H.~Tanaka for useful discussions on the subject of this paper.
\end{acknowledgments}

\end{document}